\documentclass[aps,prl,reprint,english]{revtex4-1}
\usepackage{fontenc}
\usepackage[latin1]{inputenc}
\usepackage{color}
\usepackage{graphicx}
\usepackage{amssymb}
\usepackage{babel}
\usepackage{appendix}
\usepackage{amsmath}

\begin{document}
\title{Analytic theory of optical nano-plasmonic metamaterials}
\author{Angela Demetriadou}%
\email{a.demetriadou06@imperial.ac.uk}
\author{Ortwin Hess}%
\email{o.hess@imperial.ac.uk}
\affiliation{Blackett Laboratory, Department of Physics, Imperial College London, London, SW7 2AZ, United Kingdom}%

\date{\today}%
\revised{}%

\begin{abstract}
Recent advances in nano-fabrication techniques allow for the manufacture of optical metamaterials, bringing their unique and extra-ordinary properties to the visible regime and beyond. However, an analytical description of optical nano-plasmonic metamaterials is challenging due to the characteristic optical behaviour of metals. Here we present an analytical theory that allows to bring established microwave metamaterials models to optical wavelengths. This method is implemented for nano-scaled plasmonic wire-mesh and tri-helical metamaterials, and we obtain an accurate prediction for their dispersive behaviour at optical and near-IR wavelengths. 
\end{abstract}

\maketitle

The metamaterial paradigm `function from structure' has in recent years opened the door to exciting new materials with unprecedented properties such as negative refractive index~\cite{pendry2000}. Moreover, in combination with transformation optics, it allowed a conception of dramatic new functionalities such as perfect lensing~\cite{pendry2000},  cloaking~\cite{pendry:science2006,schurig:science2006} or broadband slow and stopped light~\cite{tsakmakidis:nature2007}. As the meta-atoms that make up a metamaterial characteristically consist of metal structures that are smaller than the wavelength,  metamaterial principles such as invisibility cloaking~\cite{pendry:science2006,schurig:science2006} were first experimentally demonstrated in the microwave regime. Recent advances in nano-fabrication methods, particularly bottom-up and self-assembly techniques now allow for the design and manufacture of artificial nano-scaled metallic structures with  complex structural features~\cite{wegener2009,vukovic:acsnano2011,vignolini:advmaterials2011} forming nano-scaled metamaterials operating at optical (near infrared and visible) frequencies. 

In the conception of metamaterials, analytical theories have been the trailblazers for their design and opened our understanding of their physical properties. For chiral metamaterials this has been particularly successful in the microwave region~\cite{demetriadou:iopcondmatter2009,demetriadou:physicaB2010,demetriadou:njp2012}. At optical wavelengths, the forces excerted by incident lightfields on the nearly-free (conduction) electrons of a metal lead to collective electron-photon oscillations that typically occur on scales of just a few tens of nanometres or less. These collective electron-photon oscillations can either be localised (localised surface plasmons, LSPs) or propagating (surface plasmon polaritons, SPPs), forming areas of high intensity around and inside the plasmonic nanoparticles, the meta-atoms. In addition to nano-localisation of incident light-waves, surface plasmons may efficiently allow for the reverse effect, i.e. the out-coupling to the far-field continuum of the near-field of emissive molecules and quantum dots placed adjacently to the metal metamolecule, thereby acting as nanoantennas.  Clearly, analytical metamaterials models that are very appropriate for microwaves fall short of describing the optical response of nano-scaled metallic structures and nanoplasmonic metamaterials. Indeed, it was found through experimental results and numerical calculations~\cite{hur:angewchem2011,sangsoon:2013} that the actual performance of these structures is significantly red-shifted compared to what one may have expected from their geometrical parameters~\cite{novotny:prl2007}. 
\begin{figure}
\begin{centering}
\includegraphics[scale=0.33]{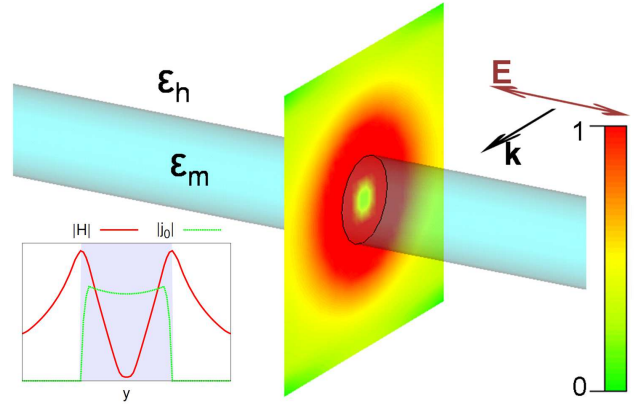}
\caption{A nano-wire made from metal of permittivity $\varepsilon_{m}$ and surrounded by a host medium of $\varepsilon_{h}$ is completely penetrated by an incident electric field $E$. The currents induced inside the wire give rise to a magnetic field $|H|$ shown at the cross-section surface of the wire. The amplitude of $|H|$ and the current density $|j_{0}|$ inside the wire are also displayed along a line parallel to the y-axis and passing though the centre of the wire (blue-shaded area represents the metallic wire). \label{fig:wire-waveguide}}
\end{centering}
\end{figure}

In this Letter, we develop an analytic theory for  the optical properties of metal nano-wires and propose a robust self-consistent method that can be used to bring well-established analytical metamaterials models from the microwave domain to optical wavelengths. In particular, we will demonstrate the significance of our theory in bringing the celebrated `wire-mesh' metamaterial model~\cite{pendry:physcondmatter1998,demetriadou:iopcondmatter2008} to the optical frequency domain with a nano-scale wire-mesh and introduce an analytical model for  tri-helical nanoplasmonic metamaterials. 

Let us start by focussing on a thin straight and infinitely extended metal nano-wire of radius $r_{w}$ and permittivity $\varepsilon_{m}$ that is embedded in a dielectric host medium with permittivity $\varepsilon_{h}$, as shown in figure~\ref{fig:wire-waveguide}. The skin-depth $\delta$ is dependent on the properties of the metal and given by: 
$\delta=\sqrt{\left((\varepsilon_{m}''-i(1-\varepsilon_{m}'))/((\varepsilon_{m}'')^2+(1-\varepsilon_{m}')^2)\right)2/(k_{0}^2\mu_{m})} $, where $k_{0}$ is the wavenumber in free space and $\mu_{m}$ the magnetic permeability of the metal.  
At microwave frequencies, any current induced on a metallic wire will flow on its surface, producing a magnetic effect around the wire and there would be no field inside the metal. Now, at optical regimes, the fields penetrate in the metal (since $r_{w} < \delta$) and they induce an electron flow throughout the metallic nano-wire and not just on its surface. The current flowing at the centre of the nanowire induces a magnetic field in a loop endorsing the current, as shown in figure~\ref{fig:wire-waveguide}. Of course, the magnitude of the induced magnetic field increases with the amount of metal it encloses and therefore it reaches a maximum value just outside the metallic nano-wire. Now these fields inside the metal slow down the effective/overall electron flow, which is modelled in this paper with the wavenumber $\gamma$ and significantly red-shift the effective response of the nano-wire.

Hence,the wavevector describing the electron flow is $k_{0}\sqrt{\varepsilon_{m}}$ and the effect of plasmon polarities (i.e. charge distribution inside the metal) is taken on board the wavevector $\gamma$, such that the overall field propagation inside the wire is governed by  $\kappa_{1}=\sqrt{k_{0}^{2}\varepsilon_{m}-\gamma^2}$\cite{novotny:prl2007}. The fields in the dielectric host medium are affected in a similar way by the plasmonic excitation, and therefore we have $\kappa_{2}=\sqrt{k_{0}^{2}\varepsilon_{h}-\gamma^2}$. Hence, $\gamma$ dominates the red-shift to the dispersive behaviour of nano-metallic structures from the $\omega$ frequency predicted from microwave analytical models, to $\omega'$ that is actually observed from experimental results and numerical calculations. 

\begin{figure}
\begin{centering}
\includegraphics[scale=0.65]{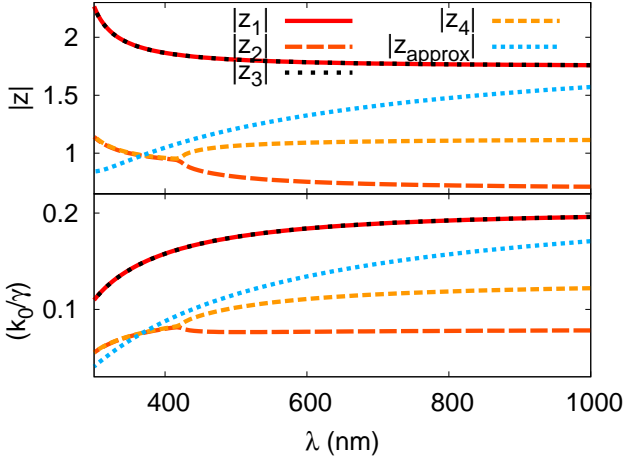}
\caption{The four solution ($|z_{i}|$) plotted for Drude-silver metal, with the approximated solution  ($|z_{approx}|$) of Novotny, which ignores metallic losses. \label{fig:z-plot}  }
\end{centering}
\end{figure}

The relationship between the two frequencies ($\omega$ and $\omega'$) is therefore dependent on $\gamma$ as~\cite{novotny:prl2007}: 
\begin{equation}
\omega=\frac{\omega'}{\left(\frac{k_{0}}{\gamma}\right)-\frac{4r_{w}\omega'}{2\pi c_{0}}} \label{eq:frequency-convertion}
\end{equation}
where the only unknown parameter is $\gamma$. Here, we find an exact solution for $\left(k_{0}/\gamma\right)$ and then we replace all $\omega$ parameters in established microwave analytical models for wire-mesh and tri-helical metamaterials with~\eqref{eq:frequency-convertion}, to obtain their optical dispersive behaviour.

In order to find $\gamma$, we observe the field behaviour in figure~\ref{fig:wire-waveguide}, which is identical to the $TM_{0}$ mode field profile for a cylindrical dielectric waveguide. The nano-wire thus effectively acts as a cylindrical dielectric waveguide since it allows the field to completely penetrate it. Hence, we use the waveguide theory to model the field penetration and plasmonic excitation for a metal nano-wire and we find $\gamma$ by solving~\cite{collin:1991}:
\begin{equation}
\frac{\varepsilon_{m}}{\kappa_{1}r_{w}}\frac{J_{1}(\kappa_{1}r_{w})}{J_{0}(\kappa_{1}r_{w})}=\frac{\varepsilon_{h}}{\kappa_{2}r_{w}}\frac{H_{1}^{(1)}(\kappa_{2}r_{w})}{H_{0}^{1}(\kappa_{2}r_{w})} \label{eq:waveguide}
\end{equation}
where $J_{n}$ and $H_{n}^{(1)}$ are the cylindrical Bessel and Hankel functions of the first kind.
\begin{figure}
\begin{centering}
\includegraphics[scale=0.65]{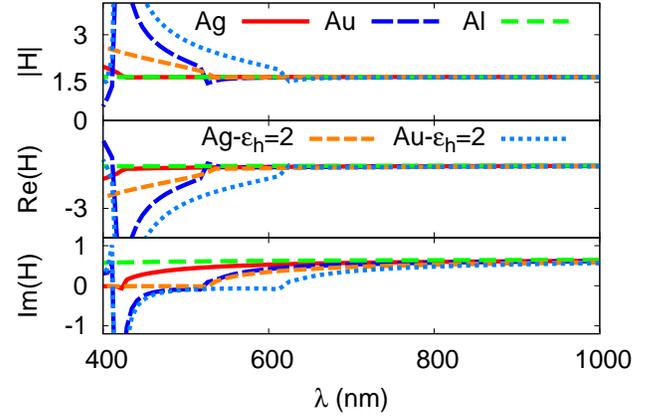}
\caption{The H-function against the frequency for silver-Ag and gold-Au($\omega_{mp}=1.3544e16$(rad/s), $g=1.5367e14$(1/s), $\varepsilon_{\infty}=9.0685$) in vacuum and $\varepsilon_{h}=2$, and Aluminium-Al ($\omega_{mp}=2.295e16$(rad/s), $g=1.27e14$(1/s), $\varepsilon_{\infty}=0.8476$). \label{fig:H}}
\end{centering}
\end{figure}
For thin wires (i.e. $\lambda\gg r_{w}$) the Bessel and Hankel functions are expanded to first order in $\kappa r_{w}$ terms and the above equation can be reduced to~\cite{novotny:prl2007}:
\begin{equation}
z^{2}+a(z)+\frac{2 \varepsilon_{h}}{\varepsilon_{m}}\left(\frac{2a(z)-2a(z)^{2}-1}{4a(z)}\right)=0 \label{eq:z-equation}
\end{equation}
where $a(z)=\Gamma-\ln(2)+\ln|z|$, $\Gamma=0.577$ is the Euler constant. We approximate $\ln|z|\sim 2\left(\frac{z-1}{z+1}\right)$, which is valid for $0.5<|z|<3$. Therefore, $a(z)\simeq a_{0}+2\left(\frac{z-1}{z+1}\right)$, where $a_{0}=\Gamma-\ln(2)$. Also, where $z$~\cite{novotny:prl2007}:
\begin{equation}
z=\sqrt{\frac{2\varepsilon_{h}}{-\varepsilon_{m}}}\left(\frac{k_{0}/\gamma}{k_{0}r_{w}\sqrt{1-\varepsilon_{h}(k_{0}/\gamma)^{2}}}\right)=\sqrt{\frac{2\varepsilon_{h}}{-\varepsilon_{m}}}\widetilde{z} \label{eq:z-definition}
\end{equation}
There are four solutions to~\eqref{eq:z-equation}, plotted in figure~\ref{fig:z-plot} for silver, which have a rather long and complex format:
\begin{eqnarray}
z_{1,2}&=&-\frac{a_{0}}{2(2+a_{0})}- \frac{1}{2} \sqrt{A} \mp \frac{1}{2}\sqrt{B-C}	 \label{eq:z-solutions} \nonumber \\
z_{3,4}&=&-\frac{a_{0}}{2(2+a_{0})}+ \frac{1}{2} \sqrt{A} \mp \frac{1}{2}\sqrt{B+C} 
\end{eqnarray}
where 
\begin{eqnarray}
A&=&\left(\frac{a_{0}}{2+a_{0}}\right)^{2} -\frac{J}{2(2+a_{0})\varepsilon_{m}}+H\\
B&=&2\left(\frac{a_{0}}{2+a_{0}}\right)^{2} -\frac{J}{2(2+a_{0})\varepsilon_{m}}-H \\
C&=&-8\left(\frac{a_{0}}{2+a_{0}}\right)\left(A-H+\frac{G}{a_{0}\varepsilon_{m}}\right)
\end{eqnarray}
and $J=3\varepsilon_{m}(2+5a_{0}+a_{0}^{2})-\varepsilon_{h}(5+6a_{0}+2a_{0}^{2})$, $G=2\varepsilon_{m}(a_{0}+2)(a_{0}-2)-\varepsilon_{h}(2a_{0}^2-2a_{0}-7)$ and $H$ is a complex functions of $\varepsilon_{h}$ and $\varepsilon_{m}$, whose full format is shown in the supplementary material.
In figure~\ref{fig:H}, function $H$ is plotted for different metals and host media. 
We observe that for frequencies where the real part of the metal's permittivity is $\left(\varepsilon_{m}/\varepsilon_{h}\right)<-5$, $H$ is independent of the metal's and host medium's material properties. Therefore, to slightly simplify solutions in~\eqref{eq:z-solutions}, $H$ can safely be approximated by $H=-1.537+i 0.658712$ for wavelengths where $\left(\varepsilon_{m}/\varepsilon_{h}\right)<-5$. The four solutions of $|z|$ are plotted in figure~\ref{fig:z-plot} for a Drude-like silver of $\varepsilon_{\infty}=4.028$, plasma frequency $\omega_{mp}=1.39e16$~(rad/s) and collision frequency $g=3.14e13$(1/s). From figure~\ref{fig:z-plot}, one can see that $|z_{1}|$ and $|z_{3}|$ are always degenerate. On the other hand, $|z_{2}|$ and $|z_{4}|$ are only degenerate for wavelengths where $\left(\varepsilon_{m}/\varepsilon_{h}\right)>-5$. Finally, we note that all our solutions are within the approximation limit $0.5<|z|<3$ that we took earlier for $ln|z|$.

Using equations~\eqref{eq:frequency-convertion} and~\eqref{eq:z-definition}, we can now obtain $\left(k_{0}/\gamma\right)$ that is also  plotted in figure~\ref{fig:z-plot} for all four $z$ solutions. Finally, the frequency shift occurring in dispersive metallic media, when they are completely penetrated by electromagnetic waves is given by:
\begin{equation}
\omega=\frac{\omega'}{\sqrt{\frac{\left(\omega'r_{w}\widetilde{z}\right)^{2}}{c_{0}^{2}+\varepsilon_{h}\left(\omega'r_{w}\widetilde{z}\right)^{2}}}-\frac{4r_{w}\omega'}{2\pi c_{0}}} \label{eq:w-PEC}
\end{equation}
where $\omega$ is the frequency found in any microwave analytical dispersive model and $\omega'$ is the new frequency accounting for the field penetration and plasmonic excitation of metals at optical frequencies.  Note that $\widetilde{z}$ is dispersive (i.e.  $\widetilde{z}(\omega')$) since it is a function of $\varepsilon_{m}$. Equation~\eqref{eq:w-PEC} can now be implemented easily into any established microwave analytical metamaterial model, bringing them to optical wavelengths. Finally,~\eqref{eq:w-PEC} is valid for metal nano-structures of circular cross-section. If one wishes to obtain the relationship between $\omega$ and $\omega'$ for metal nano-structures of square or orthogonal cross-sections, then~\eqref{eq:waveguide} needs to be rewritten accordingly (and follow the same methodology). In this article, we choose to derive~\eqref{eq:w-PEC} for circular cross-sections, since at the nano-metre scale metal structures most commonly have approximately a circular-like cross-section.

Novotny in~\cite{novotny:prl2007}, derived an approximate solution for~\eqref{eq:z-solutions} assuming loss-less Drude metals. His approximate solution  $\widetilde{z}_{approx}=\alpha_{1}+\alpha_{2}\frac{\omega_{pm}}{\omega}$, also plotted in figure~\ref{fig:z-plot} (blue dotted line) with our exact solutions. The approximation provides an indication for the order of magnitude of the frequency shift for simple metallic designs, such as a dipole optical antenna as it was intended to do (where $\omega_{pm}$ is the plasma frequency of the metal and $\alpha_{1}$ and $\alpha_{2}$ are coefficients given in~\cite{novotny:prl2007}). 

We note that despite the fact that Novotny's approximate solution \cite{novotny:prl2007} gives quite reasonable results for simple systems such as optical dipole antennas (rods of short length), it cannot predict the properties of complex metallic geometries. In particular, metamaterials that quite often are complex designs of several wavelengths in size (i.e. wire-mesh, tri-helical and gyroidal metamaterials). Furthermore, losses are an important aspect for metamaterial performance and cannot be neglected \cite{hess:naturematerials2012}. In contrary to $\widetilde{z}_{approx}$, our exact solutions in~\eqref{eq:w-PEC} accounts for metallic losses that can significantly affect the performance of optical metamaterials. However, the most important advantage of our model is that it allows for any metal (not just Drude-metals) to be considered and in fact experimental data (in the form of a fitting function) to be used for $\varepsilon_{m}$ to even account for inter sub-band transitions. 
Additionally, the model described in this paper can be also applied for both inhomogeneous and non-linear host media (i.e. gain and kerr media), by simply expressing $\varepsilon_{h}$ with the appropriate function.
The derived analytical formulation in this paper unveils the physics governing the electromagnetic interaction of metal nano-structures with light in the infra-red and visible regimes and consequently, modelling the homogenized electromagnetic behaviour of optical metamaterials.

To demonstrate the generality of our theory on analytic methods, we now apply it to two different nanoplasmonic metamaterials, the wire-mesh~\cite{pendry:physcondmatter1998,belov:physreviewb2003,demetriadou:iopcondmatter2008} and the tri-helical metamaterial (THM)~\cite{demetriadou:njp2012}.

\begin{figure}
\begin{centering}
\includegraphics[scale=0.65]{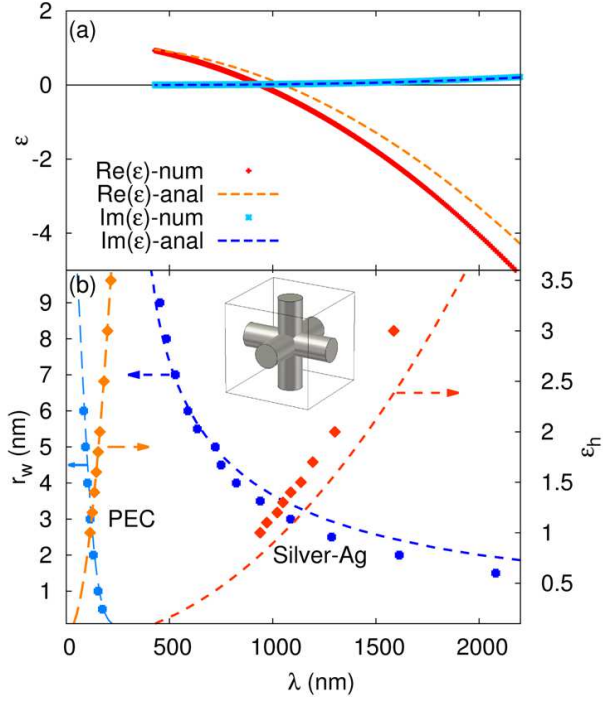}
\caption{ (a)The analytical prediction (lines) and numerical calculations (points) for a silver wire-mesh with $r_{w}=3.5nm$ and periodicity $a=40nm$. (b) The cut-off frequency variation for different $r_{w}$ and $\varepsilon_{h}$ of perfect electric conductor (PEC) and silver wire-mesh nano-metamaterial. \label{fig:wire-mesh-lp}}
\end{centering}
\end{figure}
The classic wire-mesh metamaterial is composed of thin wires aligned with the three orthogonal axes (as shown in the inset of figure~\ref{fig:wire-mesh-lp}(b)) and behaves as an artificial plasma~\cite{pendry:physcondmatter1998,belov:physreviewb2003,demetriadou:iopcondmatter2008}. Its effective electric permittivity obeys the Drude model:
\begin{equation}
\varepsilon_{wm}=1-\frac{\omega_{p-wm}^{2}}{\omega^{2}+i \gamma \omega} \label{eq:wire-mesh}
\end{equation}
where $\omega_{p-mp}$ is the cut-off frequency given by~\cite{pendry:physcondmatter1998,belov:physreviewb2003,demetriadou:iopcondmatter2008}:
$\omega_{p-wm}=c_{0}\sqrt{2 \pi/(a^{2}(\ln\left(\sqrt{\frac{a^{2}}{\pi r_{w}^{2}}}\right)+0.5275))}$.
Note that we omitted $\varepsilon_{h}$ in the above formula, since it is taken into account from our model in~\eqref{eq:w-PEC}.
Now, if $r_{w}\leq \delta$, we substitute $\omega$ in~\eqref{eq:wire-mesh} with~\eqref{eq:w-PEC}. Assuming silver nano-wires of radius $r_{w}=3.5nm$ arranged in a periodic square lattice of $a=40nm$, we calculate $\varepsilon_{wm}$ using~\eqref{eq:wire-mesh} and~\eqref{eq:w-PEC} and is plotted in figure~\ref{fig:wire-mesh-lp}(a) with numerical calculations~\footnote{Simulation calculations were performed using finite-integration techniques (CST Microwave Studio)} for comparison. The numerical values were retrieved from S-parameter calculations, using the method described in~\cite{smith:pre2005}. We see a good agreement between the two over a large wavelength spectrum. It can be shown that the small discrepancies for higher $\varepsilon_{h}$ are due to the initial analytical model in~\eqref{eq:wire-mesh}, and scale with the wavelength.
Note that for $\widetilde{z}$ in~\eqref{eq:w-PEC}, we used solution $z_{4}$, since $z_{1}$, $z_{2}$ and $z_{3}$ gave unphysical electromagnetic properties for the optical wire-mesh metamaterial.
In figure~\ref{fig:wire-mesh-lp}(b), we plot the analytical prediction (lines) of the wavelength where $\varepsilon_{wm}=0$ for various nano-wire radii ($r_{w}$) in different host media ($\varepsilon_{h}$) that show remarkable agreement with numerical calculations (points).
It is noteworthy that the wire-mesh cut-off frequency of extremely thin wires significantly shifts from few tens of nanometre for PEC metal to few micrometres for silver. Also, small variations on the optical wire radius produce extreme shifts for the cut-off frequency and our model predicts very accurately this dramatic behaviour.
\begin{figure}
\begin{centering}
\includegraphics[scale=0.65]{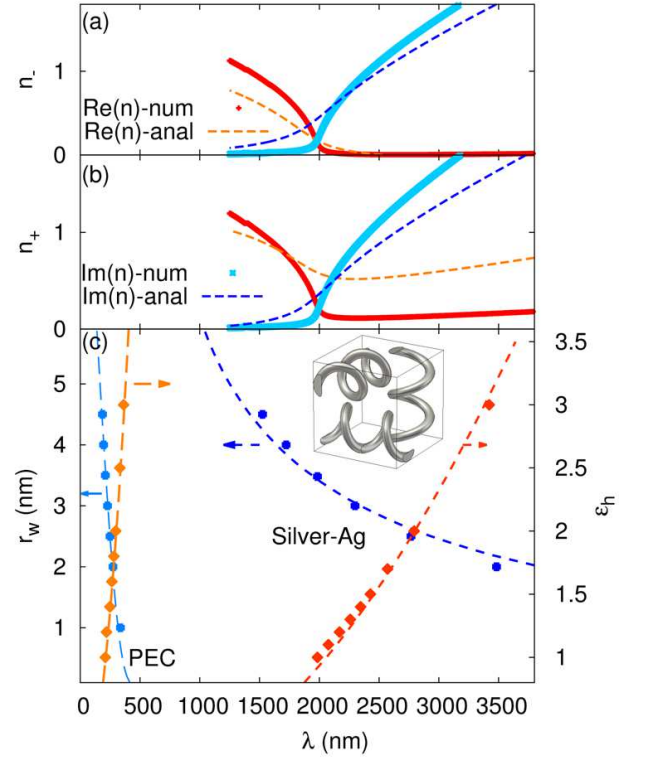}
\caption{ (a) The $n_{-}$ and (b) $n_{+}$ of a THM metamaterials of $r_{w}=3.48nm$, pitch $p=20nm$, helix radius $R=10nm$ and periodicity $a=40nm$. (c) The cut-off frequency variation for different $r_{w}$ and $\varepsilon_{h}$ of perfect electric conductor (PEC) and silver wire-mesh nano-metamaterial. \label{fig:THM-lp-r}}
\end{centering}
\end{figure}

To further demonstrate the applicability of our method on any nanoplasmonic metamaterial, we now apply it to a chiral metamaterial. The tri-helical metamaterial (THM) was introduced in~\cite{demetriadou:njp2012}, as an analytic model for a complex nanoscale self-assembled metamaterial, such as the metallic gyroids~\cite{hur:angewchem2011}. The THM is composed of three wire helices oriented along the three orthogonal axes (as shown in the inset of figure~\ref{fig:THM-lp-r}(c)). Hence, as for the wire-mesh metamaterial, the THM shows a  Drude behaviour for the effective electric permittivity:
\begin{equation}
\chi_{EE}^{-1}=\left(\frac{a}{l}\right)\left(\frac{\omega^2+i\Gamma \omega}{\omega^2+i\Gamma \omega -\omega_{p}^{2}}\right) \label{eq:chi_EE}
\end{equation}
where $\omega_{p}=c_{0}\sqrt{p/(\pi R^2 a^2 (1-\pi R^2/a^2-L p/ (2\pi R)))}$, $\Gamma$ is a loss parameter and $L$ is the self-inductance of the helical wires (see supplemental material and~\cite{demetriadou:njp2012} for more details). Unlike the wire-mesh, the effective magnetic response ($\chi_{HH}$) of the THM is resonant and not always equal to one. However, the most important characteristic of the THM is that it exhibits chiral components ($\kappa_{EH}$ and $\kappa_{HE}$), which in this case $\kappa_{EH}=-\kappa_{HE}$. Due to chirality, left-handed and right-handed circularly polarized waves propagating in a THM will experience different refractive indices, given by $n_{\pm}=1/\left(\sqrt{\chi_{EE}^{-1}(\omega)\chi_{HH}^{-1}(\omega)}\pm i\kappa_{EH}^{-1}(\omega)/c_{0}\right)$.

The electromagnetic response of a nanoscaled THM (i.e. the metallic components of THM are completely penetrated by the incident electromagnetic fields) is derived by replacing all $\omega$ terms in~\eqref{eq:chi_EE}, $\chi_{HH}$ and $\kappa_{EH}$ with~\eqref{eq:w-PEC} (see supplementary material). Again, only the $z_{4}$ solution gives physical results.
In figure~\ref{fig:THM-lp-r}(a)-(b), we plot the analytical predictions for $n_{-}$ and $n_{+}$ of a silver THM (with dimensions $r_{w}=3.48nm$, $p=20nm$,$R=6.52nm$ and $a=40nm$), along with numerical calculations that show remarkable agreement. The numerical calculation values were retrieved from S-parameter calculations, but using a different method as described in~\cite{zhao:opticsexpress2010}, due to the presence of chirality.  As before, we change the geometrical parameters of the THM by varying the wire's radius ($r_{w}$) and the host medium ($\varepsilon_{h}$). The wavelength where $\chi_{EE}=0$ for different structures is plotted in figure~\ref{fig:THM-lp-r}(c), where again the analytical (lines) and numerical (points) calculations show notable agreement. Remarkably, as for the wire-mesh metamaterial, we observe a dramatic wavelength shift between the optical and microwave models for extremely small geometrical changes in the structure.

In summary, we have derived an analytical theory for nanoplasmonic metamaterials that accounts for the field penetration and plasmonic excitation of nano-scaled metallic geometries. This method opens a door to transfer established microwave analytical models to optical wavelengths. 
In fact, our theory accounts for metallic losses, which dominate the behaviour of optical metamaterials, any metallic dispersive behaviour and  even inter sub-band transitions through the use of experimental data for the permittivity of metals.
This method is a valuable tool to analytically predict the optical behaviour of any metamaterial and coupled optical nano-antennas, but most importantly complex self-assembled nanoplasmonic metamaterials, such as gyroidal structures.

\bibliographystyle{unsrt}
\bibliography{paper}

\end{document}